\documentstyle[aps,prl,multicol,psfig]{revtex}
\newcommand{\be}{\begin{equation}}
\newcommand{\ee}{\end{equation}}
\newcommand{\con}{^{\mbox{\scriptsize{(want)}}}}
\newcommand{\pro}{^{\mbox{\scriptsize{(prod.)}}}}
\newcommand{\tra}{^{\mbox{\scriptsize{(trad.)}}}}

\newcommand{\lef}{^{\mbox{\scriptsize{(left)}}}}
\newcommand{\rig}{^{\mbox{\scriptsize{(right)}}}}
\begin{document}
\title{\bf Self-Organized Criticality in a Transient System.}

\author{Simon F. N\o rrelykke and Per Bak\thanks{Present address: 
	Department of Mathematics, Imperial College,
	London SW7 2BZ, United Kingdom}}
\address{The Niels Bohr Institute, Blegdamsvej 17, DK-2100 Copenhagen,
 Denmark.}

\date{\today}

\maketitle

\begin{abstract}
A simple model economy with locally interacting producers 
and consumers is introduced. 
When driven by extremal dynamics, 
the model self-organizes {\em not\/} to an attractor state,
but to an asymptote, on which the economy
has a constant rate of deflation, is critical, and
exhibits avalanches of activity
with power-law distributed sizes. 
This example demonstrates that self-organized critical behavior 
occurs in a larger class of systems than so far considered:
systems not driven to an attractive fixed point,
but, e.g., an asymptote, may also display self-organized criticality.
\end{abstract}

\begin{multicols}{2}

%%%%%%%%%%%%%%%%%%%%%%%%%%%%%%%%%%%%%%%%%%%%%%%%%%%%%%%%%%%%%%%%
\paragraph*{Introduction.}
It has been amply demonstrated by now
that some driven extended dissipative systems 
will self-organize into a complex critical state 
in which events of all sizes occur.
This phenomenon, called Self-Organized Criticality 
(SOC)~\cite{BTW},
has been invoked to explain phenomena such as the
experimentally observed behavior of flux-lines 
in high-$T_c$ super conductors~\cite{Field},
solar flares, and earthquakes~\cite{How}.
Several theoretical models that exhibit SOC have been 
constructed \cite{BTW,FSB,FF,IP},
for a recent review see \cite{Turcotte}.
In all these models the SOC state is a (statistically) 
stationary state.

Here we demonstrate by example that one may observe
SOC behavior also in systems which have no stationary
attractor state.
The example is a simple one-dimensional 
model economy driven by extremal dynamics. 
In this model, agents interact locally with each other 
through a fixed set of rules.
As in standard economic 
theory~\cite{Richter}, agents have
utility functions which they try to maximize.
But contrary to classical economic equilibrium theory, 
we have no `central agent,' `market maker,' or `auctioneer.' 
Maximization of utility functions is left to individual agents.
The agents have information only about agents  
with whom they interact directly, i.e., their nearest neighbors.

Agents are rational and never change their strategies,
only their prices and the quantities they buy and  produce.
By their transactions, agents make a profit,
positive or negative.
The agent who makes the most negative profit
then changes his price slightly,
in a manner that increases his profit.
In the next time step the agents do another round of
optimized transactions, and the agent now having
the most negative profit changes his price. 
This process in repeated ad infinitum.

After a transient period,
the system arrives in a state with long-range spatial 
correlations (power-laws) and deflation 
with constant rate.
The distribution of profits
displays a distinct threshold. 
Avalanches of causally connected price-changes 
by agents with profits below this threshold are observed.
The size distribution for avalanches
follows a power law.

%%%%%%%%%%%%%%%%%%%%%%%%%%%%%%%%%%%%%%%%%%%
\paragraph*{The Model.}
Consider $N$ agents numbered $n = 1,2,\ldots ,N$.
Agent number $n$ sells his product to agent number $n-1$
and buys the product produced by agent number $n+1$.
We assume that individual agents do not consume their own production, 
so in order to consume they must trade, 
and in order to trade they must produce.
Agent number $n$ produces a quantity $q_{n}$, of a good 
which is sold at a price $p_n$, per unit, 
to his neighbor numbered $n-1$. 
He subsequently buys and consumes the quantity $q_{n+1}$ of the good 
produced by his neighbor numbered $n+1$, 
who subsequently buys the good produced by {\em his\/} neighbor 
numbered $n+2$, 
etc., until all agents have made two transactions. 
This process is repeated, say once per day.

The goal of each agent is to maximize his utility
function
\be
  u_n = -c(q_n) + d(q_{n+1})
\label{eq:util}
\ee
while satisfying the constraint
\begin{equation}
  p_n q_n=p_{n+1} q_{n+1} \enspace.
  \label{eq:constraint}
\end{equation}
The first term, $-c$, in the utility function in Eq.~(\ref{eq:util})
represents the agent's cost, or discomfort,
connected with the production of $q_{n}$ units of the good he produces. 
This discomfort is an increasing function of $q$, 
and $c$ is convex because, say, the agent grows tired. 
The second term, $d$, is the utility of the good 
he buys from his neighbor.
Its marginal utility is a decreasing function of quantity $q$, 
so $d$ is an increasing, but concave, function.
This choice of $c$ and $d$ is common in economics;
see, e.g., \cite{Trejos}.

The constraint is also typical in economics. 
It is the simplest possible.
It expresses that the agents do not trust money;
they accept money as currency,
but do not want to possess any at the end of the day. 
There is no utility associated with its possession.
Also, of course, the agents want not to run out of money 
which would prevent them from getting any utility.

An explicit utility function 
is chosen for illustration and analysis,
\begin{equation}
  u_n = -\frac{1}{2}(q_n)^2 + 2\sqrt{q_{n+1}} \enspace.
  \label{eq:utility}
\end{equation} 
An agent knows the prices of his two 
neighbors at all times.
The amount of goods produced by the two neighbors 
is not known since, as we shall see,
this amount depends on the next nearest 
neighbors' prices, which again depends on {\em his} 
neighbors' prices, etc.
For the same reason, 
the demand for goods at a given time is not known either.
A similar model was invoked \cite{shubik} 
in order to explain the dynamic origin of the value of money.

Using his utility function and the prices he knows,
each agent plans how much to produce and how much to purchase, 
assuming that everything he produces will be sold, 
and that all he wants to purchase will be available.
The task is a simple optimization problem with solution
\be
  q_n\pro = \left(\frac{p_n}{p_{n+1}}\right)^{1/3}
  \label{eq:qn}
\ee
and 
\be
  q_{n+1}\con = \left(\frac{p_n}{p_{n+1}}\right)^{4/3}\enspace.
  \label{eq:qn+1}
\ee

We note in Eq.~(\ref{eq:qn}) and (\ref{eq:qn+1}) 
that the levels of production and intended consumption
are independent of absolute prices, as they depend only on ratios. 
All prices may be multiplied by a common factor,
and leave quantities produced and consumed unchanged.

Next, agent number $n$ implements his plan by producing the
quantity $q_n$, and setting it for sale at the price $p_n$.
However, his costumer, agent number $n-1$, has planned to buy the 
quantity $q_n\con$, and will do so, if  $q_n\con \le q_n\pro$.
If $q_n\con > q_n\pro$, agent $n-1$ buys the quantity available,
$q_n\pro$.
Thus, the traded amount is $q_n\tra = \min({q_n\pro,q_n\con})$.

At the end of the day agent $n$ has, unwillingly, made the profit
\be
  s_n = p_n q_n\tra - p_{n+1} q_{n+1}\tra \enspace.
  \label{eq:profit}
\ee
An agent may have negative profit if he does not sell
as much as he planned, i.e., if  
$q_n\tra < q_n\pro$.
An agent who in this way loses money, 
is not fulfilling the constraint Eq.~(\ref{eq:constraint}).
Neither is an agent who makes money.
The agent who loses most money reacts by changing his price, 
which is the only variable controlled by agents in this model.
For a given price, an agent's strategy is fixed, 
and the amount produced by the agent is determined by Eq.~(\ref{eq:qn}). 
As can be seen from Eqs.~(\ref{eq:qn}--\ref{eq:profit}), 
an agent with negative profit increases his profit
by lowering his price~\cite{inflation}. 

When agent $n$ lowers his price, his estimate
of how much he should optimally produce and consume also drops.
Conversely, his costumer agent $n-1$, 
raises his estimate of how much he should optimally buy.  
Agent $n$'s supplier does not change {\em his} estimates 
of how much he should produce and consume, 
hence he risks producing more than he can sell.
In this way an agent with negative profit
increases his profit by lowering his price, 
while potentially ``passing on'' the problem 
of negative profit to his supplier.

%%%%%%%%%%%%%%%%%%%%%%%%%%%%%%%%%%%%%%%%%%%%%%%%%%%%%
\paragraph*{Computer Simulation.}

In a simulation of the model, $N$ agents are initially given
random prices drawn from a uniform distribution on
the interval [1,2] \cite{interval},
relative price changes, $\eta$, 
are drawn from a uniform distribution on the interval 
[0, $\eta_{\text{max}}$].

The update scheme is:
(i) the levels of production and intended consumption
are found from Eqs.~(\ref{eq:qn}) and (\ref{eq:qn+1});
(ii) the profit of each agent is determined from 
Eq.~(\ref{eq:profit});
(iii) the agent with the lowest (most negative) profit is found, 
and given a new, lower price;
(iv) go to (i).

\begin{figure*}
\centerline{\psfig{figure=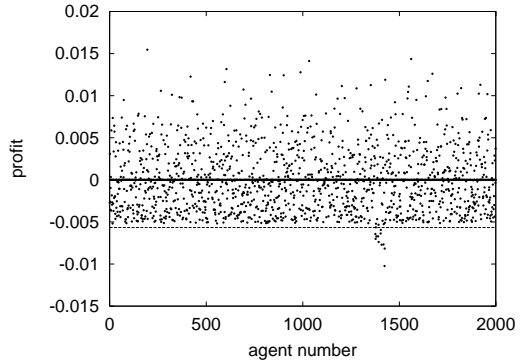,width=7cm,angle=270}}
\begin{minipage}{8cm}
\caption[]{\label{fig1}Distribution of profits after $2\cdot10^7$ time steps, 
	rescaled by 
	$\exp(2.5092 t)$; notice the delta function at zero profit.
	The dashed line marks the threshold value $f_0 = -0.0057 < f_c$.}
\end{minipage}
\end{figure*}

We studied systems of various sizes, ranging from
200 to 20,000 agents, on time-scales from
some hundreds to $10^8$ updates, and with $\eta_{\text{max}}$ ranging
from $0.1\%$ to $10\%$.
Results turned out to be insensitive to the particular value 
used for $\eta_{\text{max}}$. 
After an initial transient period, the system organized
itself into a state where the spatial distribution of
profits exhibit a clear threshold $f_c$, see Fig.~\ref{fig1}.
Few or no agents are found to have profits below this threshold,
and those found tend to be spatially located near the ``loser.''

%%%%%%%%%%

\begin{figure*}
\centerline{\psfig{figure=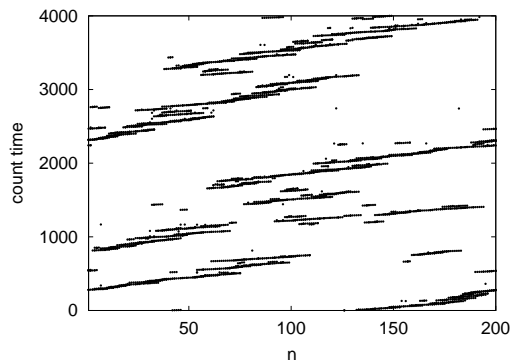,width=7cm,angle=270}}
\begin{minipage}{8cm}
\caption{\label{fig2}Spatio-temporal distribution of the losing agents 
	in an economy with  $N=200$ agents.
	Abscissa: loser's coordinate. Ordinate: time.}
\end{minipage}
\end{figure*}

Figure~\ref{fig2} shows how the loser's role 
moves through the system. 
It clearly drifts in one direction, 
because of the left-right asymmetry of the utility function.
But it also does a good deal of jumping about.

%%%%%%%%%%

The spatial correlations of the loser positions were examined
by measuring the distribution of distances between successive losers.
If the spatial jump $x$, between two successive losers was more than half
the system-size to the right, it was counted as a jump to the left.
The distribution of distances between successive losers
follow a power law distribution  asymptotically at large values
of the distances, 
i.e., the system is critical.
We fitted the distribution of distances between successive losers 
to the expression
\be
  P(x) = \mbox{A}x^{-\pi\rig} + \mbox{B}(N-x)^{-\pi\lef} + \mbox{C}
\label{eq:fit}
\enspace,
\ee
and found the exponent values 
$\pi\rig = 1.844 \pm 0.002$ and
$\pi\lef = 2.021 \pm 0.002$. 
C is a constant which takes into account the approximately
flat distribution of ``avalanche starters''~\cite{ava-starters}.
While the backing of the fit is $70\%$,
the possibility that $\pi\lef = 2$ 
cannot be ruled out~\cite{comment}.

%%%%%%%%%%
\begin{figure*}

\centerline{\psfig{figure=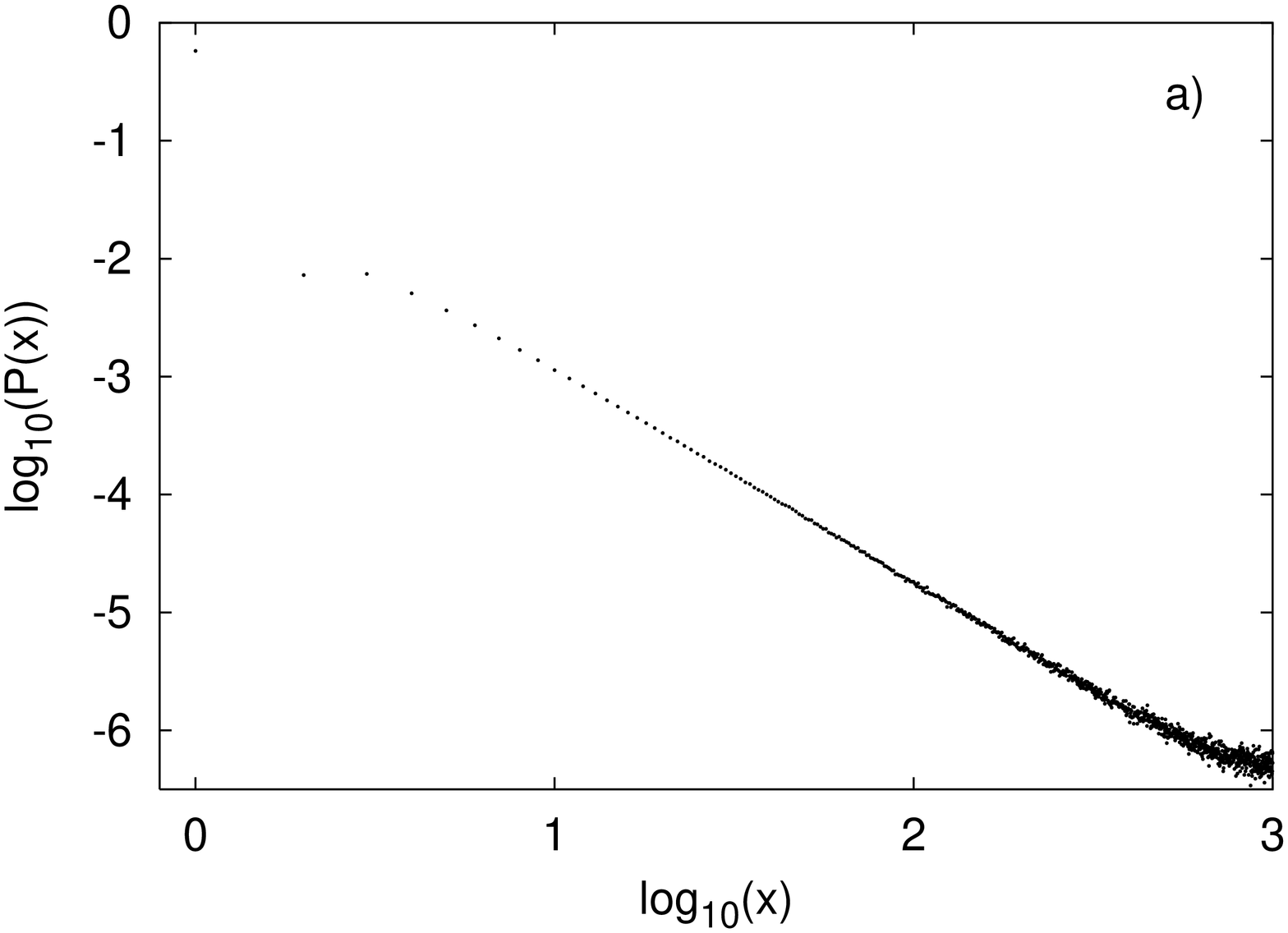,width=7cm,angle=270}}
\centerline{\psfig{figure=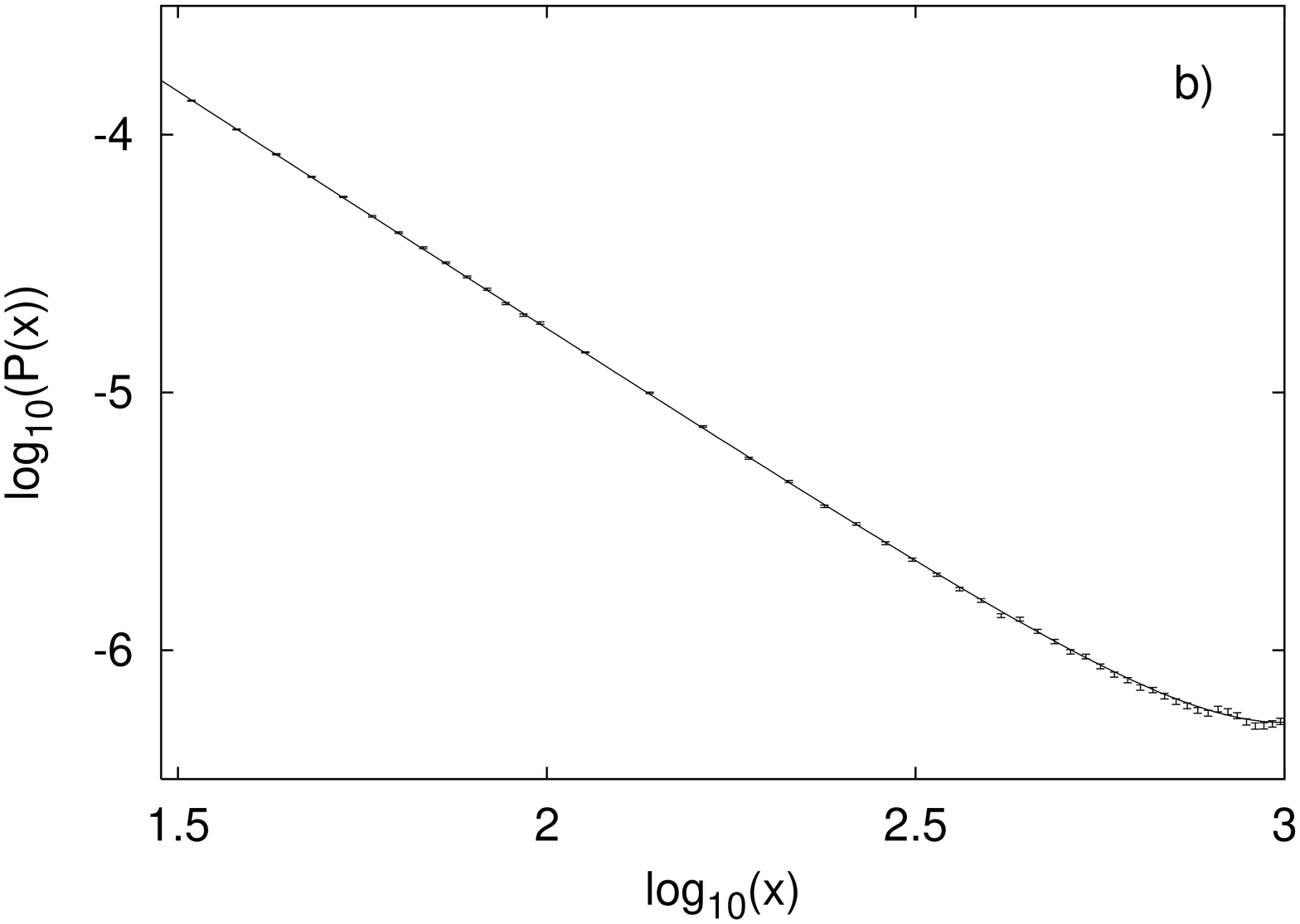,width=7cm,angle=270}}
\begin{minipage}{8cm}
\caption[]{\label{fig3}Distribution of the spatial separation
between successive losers.
Economy with 2000 agents, $\eta_{\text{max}} = 0.1\%$,
$10^8$ time steps sampled after the initial transient phase.
	(a) Jumps to the right. 
	(Jumps to the left have a similar looking distribution).
	(b) Same data as in (a), but binned. 
	Plot shows mean and RMSD of data in each bin, 
	as well as a $\chi^2$-fit of Eq.~(\ref{eq:fit})
	to the data shown in (a) and similar data for 
	jumps to the left.
	Exponents $\pi\rig=1.844\pm 0.002$ and $\pi\lef=2.021\pm 0.002$. 
	The backing of the fit is $P_{n'}(>\chi ^2) = 70\%$.
	}

\end{minipage}
\end{figure*}

Since agents keep lowering their prices, the
threshold in profit distributions decreases to zero exponentially in time,
$f_c(t) \propto \exp(-kt )$,
where
$ k  = \langle \eta \rangle/[N (1 - \langle \eta \rangle)]$
to leading order in $\langle\eta\rangle$ \cite{simeq}, 
and $\langle \cdot \rangle$ denotes an ensemble or time average.

When all profits are rescaled by $\exp(kt)$, we 
obtain stationarity of the threshold $f_c$.
We next consider the activity below a threshold 
$f_0<f_c$, and define an avalanche as the duration
of causally connected activity below this threshold. 
We refer to this duration as the avalanche size $S$.

Since it is always the agent with the lowest profit 
who changes his price and causes activity, 
it is sufficient to monitor his profit, 
and follow whether it is above or below $f_0$. 
When $s_{\mbox{\scriptsize{loser}}}>f_0$, 
all agents are above the threshold, 
and there is no active avalanche by our definition of avalanches. 
However, as the system evolves according to the update rules,
soon an agent is below threshold, and a new avalanche has been initiated.

\begin{figure*}
\centerline{\psfig{figure=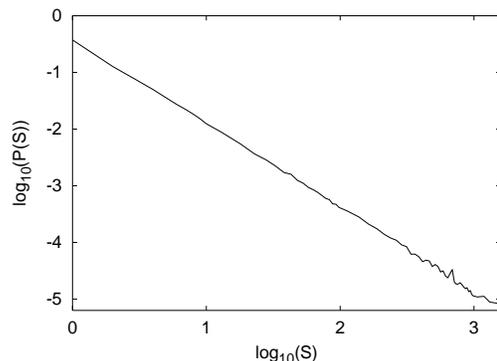,width=7cm,angle=270}}
\begin{minipage}{8cm}
\caption[]{\label{fig4}Distribution of avalanche sizes in the critical state.
	The size of an avalanche is the number of subsequent system 
	updates with (rescaled) profits less than $f_0 = -0.0057$.}
\end{minipage}
\end{figure*}

Figure~\ref{fig4} shows a log-log plot of the 
avalanche size distribution for a system of 2000 agents.
Measurements were made during $10^7$ time steps,
after discarding the first $10^6$ time steps.
We clearly see a power law $P(S) \propto S^{-1.48\pm 0.03}$.
The value of the exponent, $\tau=1.48$, is indistinguishable
from $3/2$,
the latter being the exponent of
the distribution of first-return-times 
for an unbiased random walker. 
However, when studying the avalanche size distribution for
varying positions of the threshold $f_0$,
a distribution function of a form
well known 
from percolation theory \cite{Stauffer} suggests itself\be
  P(S) = S^{-\tau} g(S(f_c - f_0)^{1/\sigma})\enspace, 
 \ee where the Fisher
exponent $\tau$, now plays the role of the avalanche
size distribution coefficient.
$g(x)$ is a scaling function, 
with the properties $g(x)\rightarrow 0$
for $x \rightarrow \infty$, 
$g(x) \rightarrow g(0)$ for $x\rightarrow 0$, 
and $\sigma$ is the
avalanche cutoff exponent \cite{Pac-Mas-Bak}.
Hence, the system cannot be adequately
described in terms of a simple unbiased random walker.

The exponent value $3/2$ is also characteristic of mean field theory, 
and was, e.g., 
obtained in the mean field treatment of the Bak-Sneppen model \cite{FSB},
which has some similarity with the model treated here.
However, as shown in \cite{BL} for the simplest possible SOC system \cite{HF},
the exponent $3/2$ can occur also in a system with fluctuations.
So one cannot from the value of our exponent conclude 
that mean field theory is exact for our model in one dimension .

%%%%%%%%%%

%%%%%%%%%%%%%%%%%%%%%%%%%%%%%%%%%%%%%%%
\paragraph*{Discussion and Conclusion.} 

Comparing to the Bak-Sneppen model \cite{FSB}, 
we use not one variable, 
but two variables, the profit and the price. 
The profits are used to find the overall
loser in the system, but we do not adjust the profit directly.
Rather, the system is driven by the losing agent's 
adjustment of his price, though he is defined by his
(lack of) profit.
Also, since $\eta_{\text{max}}$ is chosen small,
we do not randomize much.
Finally, in the Bak-Sneppen model neighbors
to a least-fit species have their fitness randomized, 
but in the present model nothing was done to the 
neighbors of a loser---only
the loser had something changed, his price.
The effect on his neighbors of this change
was predetermined and deterministic, 
and yet the system is SOC.
Thus it seems that it does not matter how a system is driven.
As long as an extremal property is chosen and adjusted in some way, 
the system will eventually build up long-range spatial correlations
and a threshold in the variable used to rank its agents.

We have shown that our model economy evolves to a critical state
when driven by extremal dynamics.
This occurs without fine-tuning of parameters, i.e.,
the system is self-organized.
We measured the distribution of spatial separations of consecutive
activity in the system,
and found two power laws (left and right) with exponents
$\pi\rig=1.844\pm 0.002$ and  $\pi\lef=2.021\pm 0.002$, with
a  70$\%$  backing of the fit. 
The system's dynamics does not have an attractive fixed point,
but only an attractive asymptote.
Hence we rescaled it to a (statistically) stationary 
state where the definition of avalanches is possible.
After this rescaling, 
we found a power law for the distribution of avalanche sizes 
with exponent $\tau = 1.48 \pm 0.03$.

The system studied here is brutally minimalistic.
There is room for several amendments towards improved realism,
with little loss in simplicity.
For example, on a two-dimensional square lattice each agent 
can have two suppliers and two costumers, 
allowing for competition, hence a market-like scenario.
In this sense, networks with higher coordination numbers
are even more realistic.
We expect criticality also in these cases,
but with different exponents.

%%%%%%%%%%%%%%%%%%%%%%%%%%%%%
\paragraph*{Acknowledgments.}
SFN thanks H. Flyvbjerg and I.~M.~Toli\'{c}
for helpful discussions. 
SFN acknowledges financial support from the L\o rup Foundation.

\bibliographystyle{abbrv}
\bibliography{references}

\begin{references}
\bibitem{inflation} This result does not depend on the specific
	choice in Eq.~(\ref{eq:utility}).
	Studying a more general version of the utility function
	$u_n = -\mbox{a}q_n^{\alpha} + \mbox{b}q_{n+1}^{\beta}$,
	we find with the specified set of rules that inflation
	will occur as long as $\alpha > \beta > 0$.
\bibitem{interval} The interval [1,2] is an arbitrary choice,
	the only demand on the initial prices is that they are positive.
\bibitem{ava-starters}  An avalanche starter is an agent with profit
	above the threshold, who is chosen as loser by our algorithm.
	This happens only when there are no agents
	with profits below the threshold,
	i.e., when an avalanche is over, per definition.
          Potentially, a new avalanche is triggered 
          by this agent.
          But it is not connected to the previous avalanche:
	Agents with profits above threshold are approximately uniformly
	distributed in space, hence so is the starting point of the new
	avalanche.
\bibitem{comment} The error we give for the fitted value of a parameter
        is just the square root of the 
	variance found for that fitting parameter.
	This is the error to be expected
	{\em when all other parameters are kept fixed}.  
	When they are not, larger errors are to be expected,
	as expressed by the variance-covariance matrix of the fit.
	If we set the value of  $\pi\lef$ to 2, i.e., fix it more than
	9 standard deviations below the fitted level, the backing drops, but
	only to $P_{1936}(>2023)=8.4\%$, i.e, not enough to 
	falsify the hypothesis that $\pi\lef = 2$.
\bibitem{simeq} 
	This relation is only a first approximation
	since the loser often has to change his price more than once in
	order to increase his profit enough. 
	A better approximation including this effect is used in the actual
	rescaling.
\end{references}

\end{multicols}

\end{document}